\documentclass[12pt]{article}

\newtheorem{theorem}{Theorem}

\newtheorem{lemma}[theorem]{Lemma}

\newtheorem{remark}[theorem]{Remark}

\input{tcilatex}

\begin{document}

\author{Yvonne Choquet-Bruhat \\
Universit\'{e} Paris 6}
\title{Non strict and strict hyperbolic systems for the Einstein equations}
\maketitle

\begin{abstract}
The integration of the Einstein equations split into the solution of
constraints on an initial space like 3 - manifold, an essentially elliptic
system, and a system which will describe the dynamical evolution, modulo a
choice of gauge. We prove in this paper that the simplest gauge choice leads
to a system which is causal, but hyperbolic non strict in the sense of Leray
- Ohya. We review some strictly hyperbolic systems obtained recently.
\end{abstract}

\section{Introduction.}

The Einstein equations equate the Ricci tensor of a pseudo riemannian
4-manifold ($V,g),$of lorentzian signature, the spacetime, with a
phenomelogical tensor which describes the sources which we take here to be
zero (vacuum case). The Einstein equations are a geometric system, invariant
by diffeomorphisms of $V$ and the associated isometries of $g.$ From the
analyst point of view they constitute a system of second order quasilinear
partial differential equations which is over determined, Cauchy data must
satisfy constraints, and underdetermined, the characteristic determinant is
identical to zero. An important problem for the study of solutions, their
physical interpretation and numerical computation is the choice of a gauge,
i.e. a priori hypothesis for instance on coordinates choice, such that the
evolution of initial data satisfying the constraints is well posed.

The geometric initial data are a 3-manifold $M$ endowed with a riemaniann
metric and a symmetric 2-tensor which will be the extrinsic curvature of $M$
embedded in $V=M\times R$. A natural gauge choice seemed to be the data on $%
V $ of the time lines by their projection $N$ on $R$ and $\beta $ on $M$.
Such a choice has been extensively used in numerical computation, though the
evolution system $R_{ij}=0$ obtained with such a choice was not known to be
well posed. In this article we will show that the system is indeed
hyperbolic in the sense of Leray-Ohya, in the Gevrey class $\gamma =2,$ and
is causal, i.e. the domain of dependence of its solutions is determined by
the light cone. When the considered evolution system is satisfied the
constraints are preserved through a symmetric first order evolution system.
Consideration of a system of order 4 obtained previously by combination of
the equations $R_{ij}=0$ with the constraints give the same result.

In section 8 we recall how the old harmonic gauge, interpreted now as
conditions on $N$ and $\beta $, gives a strictly hyperbolic evolution system
and the larger functional spaces where local existence and global
geometrical uniqueness of solutions are known.

In recent years, since the paper of C-B and York 1995, there has been a
great interest in formulating the evolution part of Eintein equations as a
first order symmetric hyperbolic system for geometrically defined unknowns.
Several such systems have been devised. Particularly interesting are those
constructed with the Weyl tensor (H.\ Friedrich, see review article 1996) or
the Riemann tensor (Anderson, C-B and York 1997) because they lead to
estimates of the geometrically defined Bel-Robinson energy used in some
global existence proofs (Christodoulou and Klainerman 1989). We recall in
the last section the symmetric hyperbolic Einstein-Bianchi system and the
corresponding Bel-Robinson energy.

\section{Einstein equations.}

The spacetime of general relativity is a pseudo riemannian manifold ($V,g),$%
of lorentzian signature (- + + +). The Einstein equations link its Ricci
tensor with a phenomelogical stress energy tensor which describes the
sources. They read

\[
Ricci(g)=\rho 
\]
that is, in local coordinates $x^{\lambda },\lambda =0,1,2,3,$ where $%
g=g_{\lambda \mu }dx^{\lambda }dx^{\mu },$

\[
R_{\alpha \beta }\stackrel{\_}{=}\frac{\partial }{\partial x^{\lambda }}%
\Gamma _{\alpha \beta }^{\lambda }-\frac{\partial }{\partial x^{\alpha }}%
\Gamma _{\beta \lambda }^{\lambda }+\Gamma _{\alpha \beta }^{\lambda }\Gamma
_{\lambda \mu }^{\mu }-\Gamma _{\alpha \mu }^{\lambda }\Gamma _{\beta
\lambda }^{\mu }=\rho _{\alpha \beta } 
\]
where the $\Gamma ^{\prime }s$ are the Christoffel symbols:

\[
\Gamma _{\alpha \beta }^{\lambda }=\frac{1}{2}g^{\lambda \mu }(\frac{%
\partial }{\partial x^{\alpha }}g_{\beta \mu }+\frac{\partial }{\partial
x^{\beta }}g_{\alpha \mu }-\frac{\partial }{\partial x^{\mu }}g_{\alpha
\beta }) 
\]
The source $\rho $ is a symmetric 2-tensor given in terms of the stress
energy tensor $T$ by 
\[
\rho _{\alpha \beta }\equiv T_{\alpha \beta }-\frac{1}{2}g_{\alpha \beta
}trT,\text{ \ \ with \ \ }trT\equiv g^{\lambda \mu }T_{\lambda \mu } 
\]

Due to the Bianchi identities the left hand side of the Einstein equations
satisfies the identities, with $\nabla _{\alpha }$ the covariant derivative
in the metric $g$ 
\[
\nabla _{\alpha }(R_{\alpha \beta }-\frac{1}{2}g_{\alpha \beta }\text{ }R)%
\stackrel{\_}{=}0,\text{ \ \ }R\equiv g^{\lambda \mu }R_{\lambda \mu } 
\]
The stress energy tensor of the sources satisfies the conservation laws
which make the equations compatible 
\[
\nabla _{\alpha }T^{\alpha \beta }=0 
\]

In vacuum the stress energy tensor is identically zero. We will consider
here only this case. The presence of sources brings up new problems specific
to various types of sources.

The Einstein equations (in vacuum) are a geometric system, invariant by
diffeomorphisms of $V$ and the associated isometries of $g.$ From the
analyst point of view they constitute a system of second order quasilinear
partial differential equations which is both undetermined (the
characteristic determinant is identical to zero) and overdetermined (one
cannot give arbitrarily Cauchy data).

\section{Intrinsic Cauchy problem}

Due to the geometric nature of Einstein's equations it is appropriate to
consider a Cauchy problem also in geometric form. The definition follows.

An initial data set is a triple $(M,\bar{g}_{0},K_{0})$ where $M$ is a 3
dimensional manifold, $\bar{g}_{0}$ a riemannian metric on $M$ and $K_{0}$ a
symmetric 2 tensor.

An extension of an initial data set is a spacetime $(V,g)$ such that there
exists an immersion $i:M\rightarrow M_{0}\subset V$ with $\stackrel{\ast }{i}%
\bar{g}_{0}$ and $i^{\ast }K_{0}$ equal respectively to the metric induced
by $g$ on $M_{0}$ and the extrinsic curvature of $M_{0}$ as submanifold of $%
(V,g).$

We say that $(V,g)$ is an einsteinian extension if $g$ satisfies the
Einstein (vacuum) equations on $V$.

A spacetime $(V,g)$ is said to be globally hyperbolic if the set of timelike
curves, between two arbitrary points is relatively compact in the Frechet
topology of curves on $V$. This definition given by Leray 1952 has been
shown by Geroch 1970 to be equivalent to the fact that ($V,g)$ possesses a
Cauchy surface, i.e. a spacelike submanifold $M_{0}$ such that each
inextendible timelike or null curve cuts $M_{0}$ exactly once.

A development is a globally hyperbolic einsteinian extension.

\section{3+1 splitting}

To link geometry with analysis one performs a 3+1 splitting of the Einstein
equations. We consider a manifold $V$ of the type $M\times R$ (the support
of a development will always be of this type). We denote by $x^{i},\in
i=1,2,3$ local coordinates in $M,$ we set $x^{0}=t\in R.$ We choose a moving
frame on $V$ such that at a point $(x,t)$ its axes $e_{i}$ coincide with the
axis of the natural frame, tangent to $M_{t}\equiv M\times \{t\},$ and its
axis $e_{0}$ is orthogonal to $M_{t},$ with associated coframe such that $%
\theta ^{0}=dt.$ A generic lorentzian metric on $V$ with the $M_{t}$ 's
spacelike reads in the associated coframe 
\[
g=-N^{2}dt^{2}+\bar{g}_{ij}\theta ^{i}\theta ^{j},\text{ \ \ with \ \ }%
\theta ^{i}\equiv dx^{i}+\beta ^{i}dt 
\]

The coefficients are time dependent geometric objects on $M.$ The scalar $N$
is called lapse, the space vector $\beta $ is called shift, $\bar{g}$ is a
riemannian metric. These elements are linked with the metric coefficients $%
g_{\alpha \beta }$ in the natural frame by the relations: 
\[
g_{ij}\equiv \bar{g}_{ij},\text{ \ \ }N^{2}=(-g^{00})^{-1},\text{ \ \ }\beta
^{i}\equiv N^{2}g^{0i} 
\]

We denote by $\bar{\nabla}$ the covariant derivative in the metric $\bar{g}.$
We have 
\[
\partial _{i}=\frac{\partial }{\partial x^{i}},\text{ \ \ }\partial
_{0}=\partial _{t}-\beta ^{i}\partial _{i},\text{ \ \ with \ \ }\partial
_{t}=\frac{\partial }{\partial t} 
\]
and we denote

\[
\hat{\partial}_{0}=\frac{\partial }{\partial t}-L_{\beta } 
\]
with $L_{\beta }$ the Lie derivative with respect to $\beta ,$ an operator
wich maps a time dependent tensor field on $M$ into another such tensor
field.

We denote by $K$ the extrinsic curvature of $M_{t}\equiv M\times \{t\}$ as
submanifold of $(V$,$g),$ i.e. we set:

\[
K_{ij}=-\frac{1}{2N}\hat{\partial}_{0}g_{ij} 
\]

A straightforward calculation (C-B 1956) gives the fundamental identities,
written in the coframe $\theta ^{0}=dt,\theta ^{i}=dx^{i}+\beta ^{i}dt,$ for
the Ricci tensor of $g:$%
\[
R_{ij}\equiv \bar{R}_{ij}-\frac{\hat{\partial}_{0}K_{ij}}{N}%
-2K_{jh}K_{i}^{h}+K_{ij}K_{h}^{h}-\frac{\bar{\nabla}_{j}\partial _{i}N}{N} 
\]
\[
R_{0i}\stackrel{\_}{=}N(-\bar{\nabla}_{h}k_{i}^{h}+\bar{\nabla}%
_{i}k_{h}^{h}) 
\]

\[
R_{00}\stackrel{\_}{=}N(\partial _{0}K_{h}^{h}-NK_{ij}K^{ij}+\bar{\nabla}%
^{i}\partial _{i}N) 
\]

\section{Constraints and evolution.}

\textbf{Constraints}

The following part of the Einstein equations do not contain second
derivatives of $g$ neither first derivatives of $K$ transversal to the
spacelike manifolds $M_{t}$. They are the constraints. They read, with $%
S_{\alpha \beta }\equiv R_{\alpha \beta }-\frac{1}{2}g_{\alpha \beta }R,$

Momentum constraint

\[
C_{i}\stackrel{\_}{=}\frac{1}{N}R_{0i}\stackrel{\_}{=}-\bar{\nabla}%
_{h}K_{i}^{h}+\bar{\nabla}_{i}K_{h}^{h}=0\text{ } 
\]

Hamiltonian constraint

\[
C_{0}\stackrel{\_}{=}\frac{2}{N^{2}}S_{00}\equiv \bar{R}%
-K_{j}^{i}K_{i}^{j}+(K_{h}^{h})^{2}=0 
\]

These constraints are transformed into a system of elliptic equations on
each submanifold $M_{t},$ in particular on $M_{0\text{ }}$ for $\stackrel{\_%
}{g}=g_{0},K=K_{0},$ by the conformal method .

\textbf{Evolution.}

The equations 
\[
R_{ij}\equiv \bar{R}_{ij}-\frac{\hat{\partial}_{0}K_{ij}}{N}%
-2K_{jh}K_{i}^{h}+K_{ij}K_{h}^{h}-\frac{\bar{\nabla}_{j}\partial _{i}N}{N}=0 
\]
together with the definition 
\[
\hat{\partial}_{0}g_{ij}=-2NK_{ij} 
\]
\ determine the derivatives transversal to $M_{t}$ of $\bar{g}$ and $K$ when
these tensors are known on $M_{t}$ as well as the lapse $N$ and shift $\beta
.$ It is natural to look at these equations as evolution equations
determining $\bar{g}$ and $K$, while $N$ and $\beta ,$ projections of the
tangent to the time line respectively on the normal and the tangent space to 
$M_{t},$ are considered as gauge variables.This point of view is conforted
by the following theorem (Anderson and York 1997, previously given for
sources in C-B and Noutchegueme 1988)

\begin{theorem}
When $R_{ij}=0$ the constraints satisfy a linear homogeneous first order
symmetric hyperbolic system, they are satisfied if satisfied initially.
\end{theorem}

Proof. When $R_{ij}=0$ we have, in the privileged frame, 
\[
R=-N^{2}R^{00} 
\]
hence 
\[
S^{00}=\frac{1}{2}R^{00}\text{ \ \ and \ }R=-2N^{2}S^{00}=2S_{0}^{0} 
\]
and 
\[
S^{ij}=-\frac{1}{2}\bar{g}^{ij}R=-\bar{g}^{ij}S_{0}^{0} 
\]
the Bianchi identities give therefore a linear homogeneous system for $%
S_{0}^{i}$ and $S_{0}^{0}$ with principal parts 
\[
N^{-2}\partial _{0}S_{0}^{i}+\bar{g}^{ij}\partial _{j}S_{0}^{0},\text{ \ and
\ \ }\partial _{0}S_{0}^{0}+\partial _{i}S_{0}^{i} 
\]
This system is symetrizable hyperbolic, it has a unique solution, zero if
the initial values are zero. The characteristics, which determine the domain
of dependence, are the light cone.

\section{Hyperbolicity non strict of $R_{ij}=0$}

An evolution part of Einstein equations should exhibit causal propagation,
i.e. with domain of dependence determined by the light cone of the spacetime
metric.

The equations $R_{ij}=0$ are, when $N$ and $\beta $ are known, a second
order differential system for $g_{ij}.$ The hyperbolicity of a quasilinear
system is defined through the linear differential operator obtained by
replacing in the coefficients the unknown by given values. In our case for
given $N,\beta $ and $g_{ij}$ the principal part of this operator acting on
a symmetric 2-tensor $\gamma _{ij}$ is 
\[
\frac{1}{2}(N^{-2}\partial _{00}^{2}-g^{hk}\partial _{hk}^{2})\gamma
_{ij}+\partial ^{k}\partial _{j}\gamma _{ik}+\partial ^{k}\partial
_{i}\gamma _{jk}-g^{hk}\partial _{i}\partial _{j}\gamma _{hk} 
\]
The characteristic matrix at a point of spacetime is the linear operator
obtained by replacing the derivation $\partial $ by a covariant vector $\xi
. $ The characteristic determinant is the determinant of this linear
operator. We take as independent unknown $\gamma _{12},\gamma _{23},\gamma
_{31,}\gamma _{11},\gamma _{22},\gamma _{33}$ and consider the 6 equations $%
R_{ij}=0$ ,same indices.

To simplify the writing we compute this matrix in a coframe orthonormal for
the given spacetime metric $(N,\beta ,g_{ij}).$ We denote by $(t,x,y,z)$ the
components of $\xi $ in such a coframe. The characteristic matrix $\mathcal{M%
}$ reads then (up to multiplication by 2):

\bigskip

$\mathcal{M\equiv } 
\begin{array}{cccccc}
t^{2}-z^{2} & zx & zy & 0 & 0 & -xy \\ 
xz & t^{2}-x^{2} & xy & -yz & 0 & 0 \\ 
yz & xy & t^{2}-y^{2} & 0 & -xz & 0 \\ 
2xy & 0 & 2zx & t^{2}-y^{2}-z^{2} & -x^{2} & -x^{2} \\ 
2xy & 2yz & 0 & -y^{2} & t^{2}-x^{2}-z^{2} & -y^{2} \\ 
0 & 2yz & 2xz & -z^{2} & -z^{2} & t^{2}-x^{2}-y^{2}
\end{array}
\allowbreak \allowbreak $

The characteristic polynomial is the determinant of this matrix. It is found
to be 
\[
Det\mathcal{M}=b^{6}a^{3},\text{ \ \ with \ \ }b=t,\text{ \ }%
a=t^{2}-x^{2}-y^{2}-z^{2} 
\]
The characteristic cone is the dual of the cone defined in the cotangent
plane by annulation of the characteristic polynomial.For our system the
characteristic cone splits into the light cone of the given spacetime metric
and the normal to its space slice. Since these charateristics appear as
multiple and the system is non diagonal it is not hyperbolic in the usual
sense. We will prove the following theorem

\begin{theorem}
When $N>0$ and $\beta $ are given, arbitrary, the system $R_{ij}=0$ is a
system hyperbolic non strict in the sense of Leray Ohya for $g_{ij}$, in the
Gevrey class $\gamma =2,$ as long as $g_{ij}$ is properly riemannian. If the
Cauchy data as well as $N$ and $\beta $ are in such a Gevrey class the
Cauchy problem has a local in time solution, with domain of dependence
determined by the light cone.
\end{theorem}

Proof.

The product $ab^{2}\mathcal{M}^{-1},$ with $\mathcal{M}^{-1}$ the inverse of
the characteristic matrix $\mathcal{M}$ is computed to be:

$
\begin{array}{cccccc}
t^{2}-x^{2}-y^{2} & -zx & -zy & 0 & 0 & xy \\ 
-zx & t^{2}-y^{2}-z^{2} & -xy & zy & 0 & 0 \\ 
-zy & -xy & t^{2}-x^{2}-z^{2} & 0 & zx & 0 \\ 
-2xy & 0 & -2zx & t^{2}-x^{2} & x^{2} & x^{2} \\ 
-2xy & -2zy & 0 & y^{2} & t^{2}-y^{2} & y^{2} \\ 
0 & -2zy & -2zx & z^{2} & z^{2} & t^{2}-z^{2}
\end{array}
\allowbreak $

We see that the elements of the matrix $ab^{2}\mathcal{M}^{-1}$ are
polynomials in $x,y,z.$ The product of this matrix by $\mathcal{M}$ is a
diagonal matrix with elements $ab^{2}$ in the diagonal$.$ Consider now the
differential operator $R_{ij}$ acting on $g_{ij}.$ Multiply it on the left
by the differential operator defined by replacing in $ab^{2}\mathcal{M}^{-1}$
the variables $x,y,z$ by the derivatives $\partial _{1},\partial
_{2},\partial _{3}.$ The resulting operator is quasi diagonal with principal
operator $\partial ^{\lambda }\partial _{\lambda }\partial _{0}^{2}.$ It is
the product of two strictly hyperbolic operators, $\partial ^{\lambda
}\partial _{\lambda }\partial _{0}$ and $\partial _{0}.$ The result follows
from the Leray-Ohya general theory.

\section{Hyperbolic non strict 4th order system.}

\begin{lemma}
The following combination of derivatives of components of the Ricci tensor
of an arbitrary spacetime : 
\[
\Lambda _{ij}\equiv \hat{\partial}_{0}\hat{\partial}_{0}R_{ij}-\hat{\partial}%
_{0}\bar{\nabla}_{(i}R_{j)0}+\bar{\nabla}_{j}\partial _{i}R_{00} 
\]
reads, when $g$ is known, as a third order quasi diagonal hyperbolic system
for the extrinsic curvature $K_{ij}.$%
\begin{eqnarray*}
\Lambda _{ij} &\equiv &\hat{\partial}_{0}\mathcal{D}K_{ij}+\hat{\partial}_{0}%
\hat{\partial}_{0}(HK_{ij}-2K_{im}K_{j}^{m})-\hat{\partial}_{0}\hat{\partial}%
_{0}(N^{-1}\bar{\nabla}_{j}\partial _{i}N)+ \\
&&\hat{\partial}_{0}(-\bar{\nabla}_{(i}(K_{j)h}\partial ^{h}N)-2N\bar{R}%
_{::ijm}^{h}K_{h}^{m}-N\bar{R}_{m(i}K_{j)}^{m}+H\bar{\nabla}_{j}\partial
_{i}N)+ \\
&&\bar{\nabla}_{j}\partial _{i}(N\bar{\Delta}N-N^{2}K.K)+\mathcal{C}_{ij}
\end{eqnarray*}
with 
\[
\mathcal{D}K_{ij}\equiv -\hat{\partial}_{0}(N^{-1}\hat{\partial}_{0}K_{ij})+%
\bar{\nabla}^{h}\bar{\nabla}_{h}(NK_{ij}),\text{ \ \ }\bar{\Delta}=\bar{%
\nabla}_{h}\bar{\nabla}^{h},\text{ \ \ }H\equiv K_{h}^{h} 
\]
and 
\[
{\mathcal{C}}_{ij}\equiv \bar{\nabla}_{j}\partial _{i}(N\partial _{0}H)-\hat{%
\partial}_{0}(N\bar{\nabla}_{j}\partial _{i}H) 
\]
\end{lemma}

Proof. A straightforward computation shows that $\mathcal{C}_{ij}$ contains
terms of at most second order in $K$ (and also in $N$) and first order in $%
\bar{g}$ (replace $\hat{\partial}_{0}g_{ij}$ by $-2NK_{ij}$). The other
terms of $\Lambda _{ij},$ except for $\hat{\partial}_{0}\mathcal{D}K_{ij}$
are second order in $K.$ All terms of $\Lambda _{ij}$ are at most second
order in $\bar{g}$ except for third order terms appearing through $\bar{%
\nabla}_{j}\partial _{i}(N\bar{\Delta}N)$. Because of these terms the system
for $\bar{g}$ and $K$ given by 
\begin{equation}
\Lambda _{ij}=0
\end{equation}
and 
\begin{equation}
\hat{\partial}_{0}g_{ij}=-2NK_{ij}
\end{equation}
is not quasi diagonal. It is not hyperbolic in the usual sense of Leray, we
will prove the following theorem.

\begin{theorem}
The system (1), (2) with unknown $\bar{g}$, $K$ is for any choice of lapse $%
N $ and shift $\beta $ equivalent to a system hyperbolic non strict in the
sense of Leray-Ohya with local existence of solutions in Gevrey classes $%
\gamma =2$ and domain of dependence determined by the light cone.
\end{theorem}

Proof. Replace in the equations $\Lambda _{ij}=0$ the tensor $K$ by $%
-(2N)^{-1}\hat{\partial}_{0}\bar{g}$: this gives a quasi diagonal system for 
$\bar{g}$, but with principal operator ($\partial _{0})^{2}\partial
^{\lambda }\partial _{\lambda }$. The result follows immediately from the
Leray-Ohya theory.

The system for $\bar{g}$, $K$ can be turned into a \emph{hyperbolic system}
by a \emph{gauge choice} as follows.

\begin{theorem}
Suppose that $N$ satisfies the wave equation. 
\begin{equation}
N^{-2}\partial _{0}\partial _{0}N-\bar{\Delta}N=f
\end{equation}
with $f$ an arbitrarily given smooth function on $M\times R.$ The system
(1),(2),(3), called $\mathcal{S},$ is equivalent to a hyperbolic Leray
system for $\bar{g},K,N$, for arbitary shift.
\end{theorem}

Proof. We use the wave equation (3) to reduce the terms $-\hat{\partial}%
_{00}^{2}(N^{-1}\bar{\nabla}_{j}\partial _{i}N)+\bar{\nabla}_{j}\partial
_{i}(N\bar{\Delta}N)$ in $\Lambda _{ij}$ to terms of third order in $N,$
second order in $g$ and $K$.

We replace the equation (3) by the also hyperbolic equation 
\begin{equation}
\hat{\partial}_{0}(N^{-2}\partial _{0}\partial _{0}N-\bar{\Delta}N)=\partial
_{0}f
\end{equation}
and replace $\hat{\partial}_{0}\bar{g}$ by $-2NK$ wherever it appears. Then
the equation (4) is third order in $N,$ while first order in $\bar{g}$ and $%
K.$ We call $\mathcal{S}^{\prime }$ the system thus modified.

We give to the equations and unknowns the Leray-Volevic indices: 
\begin{equation}
m(1)=0,\text{\ \ \ }m(2)=2,\text{ \ \ }m(3)=1
\end{equation}
\begin{equation}
n(g)=n(K)=3,\text{ \ }n(N)=4
\end{equation}
The principal matrix of the system $\mathcal{S}^{\prime }$ is then diagonal
with elements the hyperbolic operators $\partial _{0}\partial ^{\alpha
}\partial _{\alpha }$ or $\partial _{0}.$

If the equation (3) is satisfied on the initial submanifold as well as $%
\mathcal{S}^{\prime }$, the equation (3$)$ and the system $\mathcal{S}$ are
satisfied.

\begin{remark}
The system $\Lambda _{ij}=0$ has the additionnal property to satisfy a
polarized null condition, that is the quadratic form defined by the second
derivative of $A$ at some given metric $g,$ $\Lambda "_{ij}(g)(\gamma
,\gamma ),$ vanishes when $\gamma =\ell \otimes \ell $ with $\ell $ a null
vector for the spacetime metric $g$ such that $\gamma $ is in the kernel of
the first derivative of the Ricci tensor of spacetime at $g$ (C-B 2000)$.$
\end{remark}

\section{An hyperbolic second order system}

A variety of hyperbolic evolution systems for Einstein equations have been
obtained, with a speed greatly increasing in recent years, by replacing the
trivial gauge choice (which is the data of $N$ and $\beta $ on $V)$ by more
elaborate ones, together with combining the evolution equations with the
constraints. The hope in changing the gauge is to find systems either better
suited to the study of global existence problems, or more stable under
numerical codes. We give some references in the bibliography. We will return
below to the original gauge choice (C-B 1952) in the perspective of
conditions on the lapse and the shift.

The following identity was already known by De Donder, Lanczos and Darmois.
It splits the Ricci tensor with components $R_{\alpha \beta }$ in the
natural frame into a quasi linear quasi diagonal wave operator and 'gauge'
terms, as follows:

\[
R_{\alpha \beta }\stackrel{\_}{=}R_{\alpha \beta }^{(h)}+\frac{1}{2}%
(g_{\beta \lambda }\frac{\partial }{\partial x^{\alpha }}F^{\lambda
}+g_{\alpha \lambda }\frac{\partial }{\partial x^{\beta }}F^{\lambda }) 
\]
with 
\[
R_{\alpha \beta }^{(h)}\stackrel{\_}{=}-\frac{1}{2}g^{\lambda \mu }\frac{%
\partial ^{2}}{\partial x^{\lambda }\partial x^{\mu }}g_{\alpha \beta
}+H_{\alpha \beta }=0 
\]
where $H_{\alpha \beta }$ is a quadratic form in first derivatives of $g$
with coefficients polynomials in $g$ and its contravariant associate.

The $F^{\lambda }$ are given by 
\[
F^{\lambda }\equiv g^{\alpha \beta }\Gamma _{\alpha \beta }^{\lambda }\equiv
\nabla ^{\alpha }\nabla _{\alpha }x^{(\lambda )} 
\]
They are non tensorial quantities, result of the action of the wave operator
of $g$ on the coordinate functions. For this reason their vanishing is
called 'harmonicity condition'.

The contravariant components of the Ricci tensor admit an analogous
splitting, namely: 
\[
R^{\alpha \beta }\stackrel{\_}{=}R_{(h)}^{\alpha \beta }+\frac{1}{2}%
(g^{\alpha \lambda }\frac{\partial }{\partial x^{\lambda }}F^{\beta
}+g^{\beta \lambda }\frac{\partial }{\partial x^{\lambda }}F^{\alpha }) 
\]
with 
\[
R^{\alpha \beta }=\frac{1}{2}g^{\lambda \mu }\frac{\partial ^{2}}{\partial
x^{\lambda }\partial x^{\mu }}g^{\alpha \beta }+K^{\alpha \beta } 
\]
Let us denote by $R_{\alpha \beta }^{(\theta )}$ the components of the Ricci
tensor in the previous frame $\theta ^{0}=dt,$ $\theta ^{i}=dx^{i}+\beta
^{i}dt$, to distinguish them from the components in the natural frame \ now
denoted $R_{\alpha \beta }.$ These components are linked by the relations: 
\[
R_{ij}^{(\theta )}=R_{\alpha \beta }\frac{\partial (dx^{\alpha })}{\partial
\theta ^{i}}\frac{\partial (dx^{\beta })}{\partial \theta ^{j}}=R_{ij},\text{
\ \ } 
\]
\[
R^{00}=R_{(\theta )}^{\alpha \beta }\frac{\partial (dt)}{\partial \theta
^{\alpha }}\frac{\partial (dt)}{\partial \theta ^{\beta }}=R_{(\theta
)}^{00} 
\]
\[
R_{(\theta )}^{0i}=R^{\alpha \beta }\frac{\partial (\theta ^{0})}{\partial
(dx^{\alpha })}\frac{\partial (\theta ^{i})}{\partial (dx^{\beta })}%
=R^{00}+R^{0j} 
\]
The equations $R_{ij}^{(h)}=0$ are a quasidiagonal second order system for $%
g_{ij}$ when $N$ and $\beta $ are known. The equations $R_{(h)}^{00}=0$ and $%
R_{(h)}^{0i}$ are quasilinear wave equations for $N$ and $\beta $ when the $%
g_{ij}^{\prime }s$ are known: we interpret these equations as gauge
conditions. The set of all these equations constitute a quasidiagonal second
order system for $g_{ij},$ $N$ and $\beta $, hyperbolic and causal as long
as $N>0$ and $\bar{g}$ is properly riemannian.

The Bianchi identities show that for a solution of these equations the
quantities $F^{\lambda }$ satisfy a linear homogeneous quasidiagonal
hyperbolic and causal system. Its initial data can be made zero by choice of
initial coordinates if and only if the geometric initial data $\bar{g},K$
satisfy the constraints. A solution of our hyperbolic system satisfies then
the full Einstein equations.

The following local existence and uniqueness theorem improves the
differentiability obtained in the original theorem of C-B 1952 who used $%
C^{k}$ spaces and a constructive (parametrix) method. The improvement to
Sobolev spaces (one can endow $M$ with a given smooth riemannian metric to
define those spaces) with $s\geq 4$ for existence and $s\geq 5$ for
geometric uniqueness is given in C-B 1968 using Leray's results. The
improvement given in the theorem was sugested by Hawking and Ellis 1973,
proved by semigroup methods by Hughes, Kato and Marsden 1978 and by energy
methods by C-B, Christodoulou and Francaviglia 1979. The other hyperbolic
systems constructed in the past twenty years did not lead, up to now, to
further improvement on the regularity required of the Cauchy data. Such an
improvement would be an important step.

\begin{theorem}
Given an initial data set, $\stackrel{\_}{g_{0}},K_{0}\in
H_{s}^{local},H_{s-1}^{local}$ satisfying the constraints, there exists an
einsteinian extension if $s\geq 3.$
\end{theorem}

The question of uniqueness is a geometrical problem. It is in general easy
to prove that the solution is unique in the chosen gauge, for instance in
the harmonic gauge recalled above. But two isometric spacetimes must be
considered as identical. The following theorem (C-B and Geroch) gives this
geometric uniqueness (maximal means inextendible).

\begin{theorem}
The development of an initial data set is unique up to isometries in the
class of maximal developments if $s\geq 4$. The domain of dependence is
determined by the light cone of the spacetime metric.
\end{theorem}

The proof of C-B and Geroch considers smooth data and developments, the
refined \ result is due to Chrusciel 1996. The geometric uniqueness in the
case $s=3$, even the local one, is still an open problem.

\section{Bianchi equations.}

The Riemann tensor satisfies the identities 
\begin{equation}
\nabla _{\alpha }R_{\beta \gamma ,\lambda \mu }+\nabla _{\beta }R_{\gamma
\alpha ,\lambda \mu }+\nabla _{\gamma }R_{\alpha \beta ,\lambda \mu }\equiv 0
\end{equation}
it holds therefore that, modulo the symmetries of the Riemann tensor 
\begin{equation}
\nabla _{\alpha }R_{..\mu ,\beta \gamma }^{\alpha }\equiv \nabla _{\beta
}R_{\gamma \mu }-\nabla _{\gamma }R_{\beta \mu }
\end{equation}
hence if the Ricci tensor $R_{\alpha \beta }$ satisfies the vacuum Einstein
equations 
\begin{equation}
R_{\alpha \beta }=0
\end{equation}
it holds that 
\begin{equation}
\nabla _{\alpha }R_{..\mu ,\beta \gamma }^{\alpha }=0.
\end{equation}

The system (8), (11) splits as the Eintein equations into constraints,
containing no time derivatives of curvature, namely in the frame used in the
3+1\ splitting: 
\begin{equation}
\nabla _{i}R_{jk,\lambda \mu }+\nabla _{k}R_{ij,\lambda \mu }+\nabla
_{j}R_{ki,\lambda \mu }\equiv 0
\end{equation}
\begin{equation}
\nabla _{\alpha }R_{..0,\beta \gamma }^{\alpha }=0.
\end{equation}
and an evolution system 
\begin{equation}
\nabla _{0}R_{hk,\lambda \mu }+\nabla _{k}R_{0h,\lambda \mu }-\nabla
_{h}R_{0k,\lambda \mu }=0
\end{equation}
\begin{equation}
\nabla _{0}R_{:::i,\lambda \mu }^{0}+\nabla _{h}R_{:::i,\lambda \mu }^{h}=0
\end{equation}
This system has a principal matrix consisting of 6 identical 6 by 6 blocks
around the diagonal, obtained by fixing a pair $\lambda ,\mu ,$ $\lambda
<\mu .$ Each block is symmetrizable through the metric $\bar{g}$, and
hyperbolic if $\bar{g}$ is properly Riemannian and $N>0$ because the
principal matrix $M^{0}$ for the derivatives $\partial _{0}$ was, up to
product by $N^{-1}$, the unit matrix and the derivatives $\partial _{h}$ do
not contain $\partial /\partial t$.

The Bel-Robinson energy is the energy associated to this symmetric
hyperbolic system.

\begin{remark}
Following Bel one can introduce two pairs of gravitational ``electric'' and
``magnetic'' space tensors associated with the 3+1\ splitting of the
spacetime and the double two-form Riemann($g):$ 
\begin{eqnarray*}
N^{2}E_{ij} &\equiv &R_{0i,0j},\text{ \ \ }D_{ij}\equiv \frac{1}{4}\eta
_{ihk}\eta _{jlm}R^{hk,lm} \\
NH_{ij} &\equiv &\frac{1}{2}\eta _{ihk}R_{:::::,0j}^{hk},\text{ \ \ }%
NB_{ji}\equiv \frac{1}{2}\eta _{ihk}A_{0j,}^{:::::hk}
\end{eqnarray*}
where $\eta _{ijk}$ is the volume form of $\bar{g}$. The principal part of
the evolution system ressemble then to the Maxwell equations, but contains
an additional non principal part. Its explicit expression is given in
Anderson,C-B and York 1997.
\end{remark}

The Bianchi equations do not tell the whole story since they contain the
spacetime metric $g,$ which itself depends on the Riemann tensor.

A possibility to obtain a symmetric evolution system (Friedrich 1996 with
the Weyl tensor) for both $g$ and $Riemann(g)$ (Anderson, C-B and York 1997)
is to introduce again the auxiliary unknown $K$ and use $3+1$ identities,
involving now not only the Ricci tensor but also the Riemann tensor. One can
then obtain a symmetric first order hyperbolic system for $K$ and $\bar{%
\Gamma},$ the space metric connection, modulo a choice of gauge, namely the
integrated form of the harmonic time-slicing condition used before (C-B and
Ruggeri 1983). The energy associated to this system has unfortunately no
clear geometrical meaning. Determination of the metric from the Riemann
tensor through elliptic equations seems more promising for the solution of
global problems (see Christodoulou and Klainerman 1989, Andersson and
Moncrief, in preparation)

\textbf{References.}

Abrahams A., Anderson A., Choquet-Bruhat, Y., York, J.W. (1996) A non
strictly hyperbolic system for Einstein equations with arbitrary lapse and
shift. C.R.\ Acad. Sci. Paris S\'{e}rie 2b, 835-841.

Anderson A., Choquet-Bruhat, Y., York, J.W.(1997) Einstein Bianchi
hyperbolic system for general relativity. Topol. Meth. Nonlinear Anal. 10,
353-373.

Anderson A., Choquet-Bruhat, Y., York, J.W. (2000) Einstein's equations and
equivalent dynamical systems. In Cotsakis, S. (ed.) Mathematical and quantum
aspects of relativity and cosmology. Cotsakis S., Gibbons G., (ed) Lecture
Notes in Physics 537, Springer.

Andersson, L. , Moncrief, V. , private communication.

Choquet (Four\`{e}s)-Bruhat, Y. 1952\ Th\'{e}or\`{e}me d'existence pour
certains syst\`{e}mes d'\'{e}quations aux d\'{e}riv\'{e}es patielles non
lin\'{e}aires. Acta Matematica 88, 141-225.

Choquet (Four\`{e}s)-Bruhat, Y. 1956 Sur l'int\'{e}gration des \'{e}quations
de la relativit\'{e} g\'{e}n\'{e}rale.J.\ Rat. Mech. and Anal. 55, 951-966.

Choquet-Bruhat Y. 1968 Espaces temps einsteiniens g\'{e}n\'{e}raux, chocs
gravitationnels Ann. Inst. Poincar\'{e} 8, n$%
{{}^\circ}%
4$ 327-338.

Choquet-Bruhat Y. 2000 The null condition and asymptotic expansions for the
Einstein equations. Ann. der Physik 9, 258-267.

Choquet-Bruhat Y., Christodoulou D., Francaviglia M., 1978 Cauchy data on a
manifold. Ann. Inst. Poincar\'{e} A 23 241-250

Choquet-Bruhat Y., Geroch R. 1969 Global aspects of the Cauchy problem in
general relativity. Comm. Math. Phys. 14 329-335.

Choquet-Bruhat, Y., Noutchegueme, N. 1986 Syst\`{e}me hyperbolique pour les
\'{e}quations d'Einstein avec sources. C. R. Acad. Sci. Paris s\'{e}rie 303,
259-263.

Choquet-Bruhat, Y., Ruggeri T. 1983 Hyperbolicity of the 3+1 system of
Einstein equations. Comm. Math. Phys. 89, 269-275.

Choquet-Bruhat, Y., York, J.W. 1995 Geometrical well posed systems for the
Einstein equations. C.R.\ Acad. Sci. Paris s\'{e}rie 1 321, 1089-1095.

Christodoulou D., Klainerman S. 1989\ The non linear stability of Minkowski
space. Princeton University Press.

Chrusciel P. T. 1991 On the uniqueness in the large of solutions of
Eintein's equations. Proc. Cent. Math. Anal. (ANU, Canberra) 20

Friedrich, H. 1996\ Hyperbolic reductions for Einstein'equations. Class.
Quant. Grav. 13, 1451-1459.

Hawking S., Ellis G. 1973 The global structure of spacetime Cambridge Univ.
Press.

Leray J. 1953 Hyperbolic differential equations Lecture Notes, Princeton.

Leray J., Ohya Y. (1967) Equations et syst\`{e}mes non lin\'{e}aires
hyperboliques non stricts. Math. Ann. 170 167-205.

Reula O. 1998 Hyperbolic methods for Einstein's equations.

www.livingreviews.org/Articles/volume1/1998-3reula

\end{document}